\documentclass
[aps,prl,twocolumn,superscriptaddress,showpacs,floatfix]{revtex4-1}%
\usepackage{graphicx,amsmath,amssymb}
\usepackage{amsmath}
\usepackage{amsfonts}
\usepackage{amssymb}
\usepackage{graphicx}%
\providecommand{\U}[1]{\protect\rule{.1in}{.1in}}
\begin{document}
\title{Hyperfine-controlled domain-wall motion observed in real space and time}
\author{John N. Moore}
\affiliation{Department of Physics, Tohoku University, Sendai 980-8578, Japan}
\author{Junichiro Hayakawa}
\affiliation{Department of Physics, Tohoku University, Sendai 980-8578, Japan}
\author{Takaaki Mano}
\affiliation{National Institute for Materials Science, Tsukuba, Ibaraki 305-0047, Japan}
\author{Takeshi Noda}
\affiliation{National Institute for Materials Science, Tsukuba, Ibaraki 305-0047, Japan}
\author{Go Yusa}
\email{yusa@tohoku.ac.jp}
\affiliation{Department of Physics, Tohoku University, Sendai 980-8578, Japan}
\date{\today
}

\begin{abstract}
We perform real-space imaging of propagating magnetic domains in the
fractional quantum Hall system using spin-sensitive photoluminescence
microscopy. The propagation is continuous and proceeds in the direction of the
conventional current, i.e. opposite to the electron flow direction. The
mechanism of motion is shown to be connected to polarized nuclear spins around
the domain walls. The propagation velocity increases when nuclei are
depolarized, and decreases when the source-drain current generating this
nuclear polarization is increased. We discuss how these phenomena may arise
from spin interactions along the domain walls.

\end{abstract}

\maketitle

Research around magnetic domains and their dynamics has become increasingly
relevant, driven by the hunt for domain-based logic and memory
\cite{allwood,parkin}. These technologies could dramatically reduce device
heating while increasing speed. A number of scientifically innovative methods
for controlling the propagation of ferromagnetic domain walls have recently
been pioneered \cite{ramsay,miron,emori,franken,yamanouchi,yamaguchi}. Across
these works numerous interaction phenomena have been identified as driving and
assisting domain propagation, ranging from spin-transfer torques from injected
electrons and optical pulses, to torques and stabilizing influences from
Rashba fields and Dzyaloshinskii-Moriya interactions. Apart from these
interactions, another potential control parameter relevant to domain motion is
hyperfine interaction.

Coupling between conduction electrons and a material's nuclear spin bath has
been known since the 1950's to occur via hyperfine interaction \cite{feher}.
Recently, nuclear magnetic resonance (NMR) studies in semiconductors have
attracted renewed interest for their value in research on quantum information
processing, especially in quantum confined nanostructures
\cite{kronmuller99,machida,yusaN,petta,koppens,ono,yusa,latta,vink,hennel,sanada}%
. It has been reported that electron-nuclear spin coupling can lead to dynamic
nuclear polarization and sometimes can function as a good control parameter to
manipulate electron spins \cite{petta,koppens}, or may cause nuclear
polarization to act back on the electronic system in complex ways
\cite{ono,yusa,latta,vink,hennel}. One system in which hyperfine interaction
becomes relevant is the fractional quantum Hall (FQH) system
\cite{kronmuller99}, where strongly interacting electrons condense into a 2D
liquid at fractional values of the Landau level filling factor $\nu$
\cite{tsui}. At certain values of $\nu$ and magnetic field $B$, this system
plays host to a phase transition between two degenerate spin resolved
many-bodied ground states, i.e. ferromagnetic and non-magnetic states, in
which electron spin polarization $P$ is $1$ and $0$, respectively. Near this
phase transition, bringing the system out of equilibrium with a strong
source-drain current can excite into existence stripe-shaped domains, which
elongate along the Hall electric field (perpendicular to the source-drain
current direction); spin-resolved electrons passing between these domains
undergo flip-flop scattering with nuclei, producing nuclear spin polarization
$P_{\text{N}}$ near domain walls \cite{moore}.

In this Letter we investigate the hyperfine-mediated controllability of domain
walls in real space and time. Using spin-sensitive photoluminescence (PL)
microscopy, we image domains propagating through the sample in response to a
direct source-drain current $I_{\text{DC}}$. This propagation is continuous
and unidirectional. The propagation velocity increases when nuclei are
resonantly depolarized, and it shows dependencies on $\nu$ and the magnitude
of $I_{\text{DC}}$ that also suggest $P_{\text{N}}$'s tendency to reduce the
velocity. We discuss how these phenomena may arise from spin interactions
along the domain walls.

\begin{figure}[t]
\begin{center}
\includegraphics{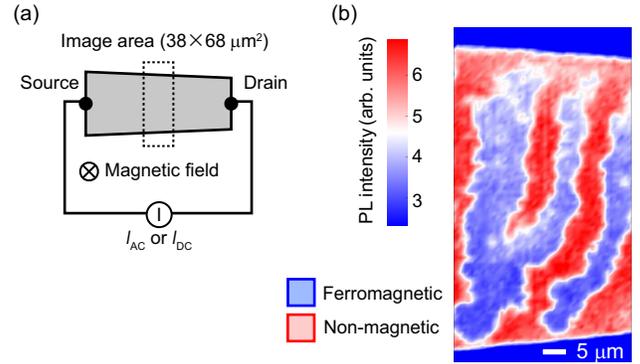}
\end{center}
\caption{(Color online) (a) Schematic of the sample. Alternating
($I_{\text{AC}}$) or direct ($I_{\text{DC}}$) current can be applied between
the source and drain. External magnetic field $B$ perpendicular to the 2D
electrons is $6.8$~T throughout. (b) $38\times68$-$\mu$m$^{2}$ spatial image
showing integrated PL intensity of charged exciton singlet peak, at
$\nu=0.666$, with $13$-Hz, $I_{\text{AC}}=60$~nA alternating source-drain
current. Image step size: $781$~nm. $T\sim60$~mK unless otherwise specified.}%
\label{fig:fig1}%
\end{figure}

\begin{figure*}[t]
\centering
\par
\begin{center}
\includegraphics{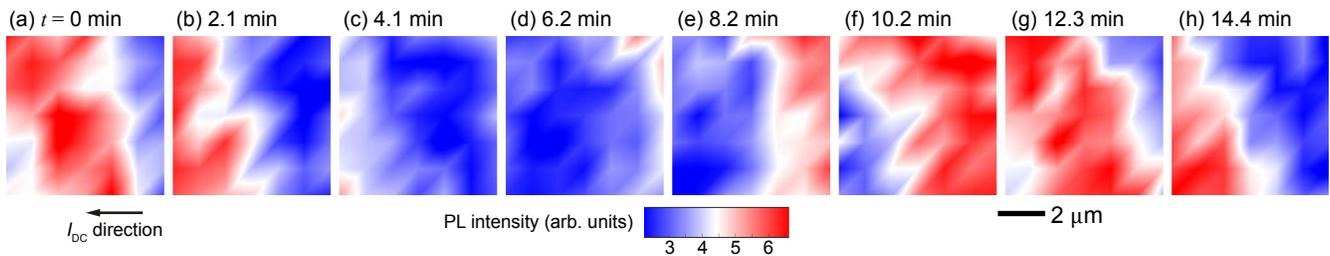}
\end{center}
\caption{(Color online) $6\times6$-$\mu$m$^{2}$ PL-intensity images of the
center of the region shown in Fig.~1(b) for $t$ of: (a) $0$; (b) $2.1$; (c)
$4.1$; (d) $6.2$; (e) $8.2$; (f) $10.2$; (g) $12.3$; and (h) $14.4$~min.
$I_{\text{DC}}=110$~nA, $\nu=0.664$. Image step size: 1~$\mu$m. See
supplementary video \cite{SI}.}%
\label{fig:fig2}%
\end{figure*}

Measurements were carried out at temperature $T\sim60$~mK in a $15$-nm-wide
GaAs/AlGaAs quantum well sample containing a FQH liquid with $\nu$ close to
$2/3$ and at the critical magnetic field ($B=6.8$~T) of the $\nu=2/3$ spin
phase transition \cite{hayakawa}. An \textit{n}-doped GaAs substrate functions
as a back gate and enables us to tune two-dimensional electron density $n_{e}$
and $\nu$ at constant $B$. The sample was measured by scanning optical
microscopy and spectroscopy. We spatially mapped the integrated PL intensity
produced by singlet-state charged excitons \cite{yusaPRL,wojs06}; this
intensity is primarily anti-correlated with the local $P$ \cite{hayakawa}, but
also is sensitive to the local $P_{\text{N}}$ \cite{moore}. Accordingly, the
non-magnetic and ferromagnetic domains are distinguished by strong and weak PL
intensities, respectively \cite{hayakawa}. For comparison to the previous
study \cite{moore}, we applied a $13$-Hz \textit{alternating }source-drain
current $I_{\text{AC}}=60$~nA to the sample near the phase transition
[Fig.~1(b)]. Upon applying the current, the striped domains that are excited
tend to be unstable for the first $\sim1$~hr; after this period they appear
static over the duration of the imaging ($\sim20$~hr). The spatial image at
$\nu=0.664$ [Fig.~1(b)] excited by $I_{\text{AC}}$ is consistent with that
reported earlier showing the formation of domain structures \cite{moore}.

When, in contrast, \textit{a direct }source-drain current $I_{\text{DC}}$ is
applied, the scenario is altered dramatically; the domains propagate spatially
[Fig.~2(a)$-$2(h); see SI video]. In Fig. 2(a), a domain wall in the right of
the image at an arbitrary time $t=0$ propagates $3-4~\mu$m to the left at
$t=4.1$~min [Fig.~2(c)]. Another domain wall propagates across the image in
the same manner [Figs.~2(d)$-$2(g)]. The propagation direction is identical to
the current direction, and reversing the current direction reverses the
propagation direction. The widths of these striped domains are consistently
preserved \cite{SI}, indicating that the domain walls all have nearly
equivalent velocities, possibly because conservation of $P$ throughout the
system is energetically favorable.

The integrated microscopic PL ($\mu$-PL) intensity obtained from a diffraction
limited spot ($\phi\sim1$~$\mu$m) at a fixed point under $I_{\text{DC}}%
=110~$nA oscillates reasonably periodically in time with a period on the order
of $\sim10$~min [Fig.~$3$(a)], suggesting that the stripe domains form with a
fairly equally spaced period. In contrast, for $I_{\text{DC}}=0~$nA, all the
imaged area contains a ferromagnetic ground state, and the PL intensity at the
fixed point is constant over time. The averaged PL intensity for
$I_{\text{DC}}=0~$nA, denoted by a dotted line in Fig. 3(a), is near the
center of oscillations observed for $I_{\text{DC}}=110~$nA. The drops in
intensity below this ferromagnetic ground-state value indicate the influence
of $P_{\text{N}}$ anti-parallel to $B$ inside of ferromagnetic phase domains.
The increases in intensity above the ground-state value are due to the combined
influence of $P_{\text{N}}$ parallel to $B$ and the vanishing of $P$ in the
non-magnetic phase domains.

In order to obtain the average domain velocity along the horizontal direction
$\upsilon_{\text{domain}}$, we measured $\mu$-PL intensities at two points
(Points~1 and 2) aligned along the length of the Hall bar [Fig.~$3$(c)]. In
this measurement, the $\mu$-PL spectrum was collected for $4$~s at Point~1;
the measurement position was then immediately shifted $4~\mu$m to the right
(Point~2) and the $\mu$-PL spectrum was again collected for $4$ s; after
returning to the original point (Point~1), the cycle was repeated. A time
delay appears in the intensity at these two points due to the domain motion,
and its average value can be determined from the cross-correlation between the
two sets of oscillations. The cross-correlation is maximum at a time lag of
$-1.330$~min [Fig.~3(d)]. The average velocity of the domains, $\upsilon
_{\text{domain}}$, at $I_{\text{DC}}=110~$nA, $\nu=0.662$, thus, is estimated
to be $\sim45~$nm/s. The strong cross-correlation confirms that the widths of
the striped domains are preserved over short distances on the order of the
domain widths.

\begin{figure*}[t]
\includegraphics{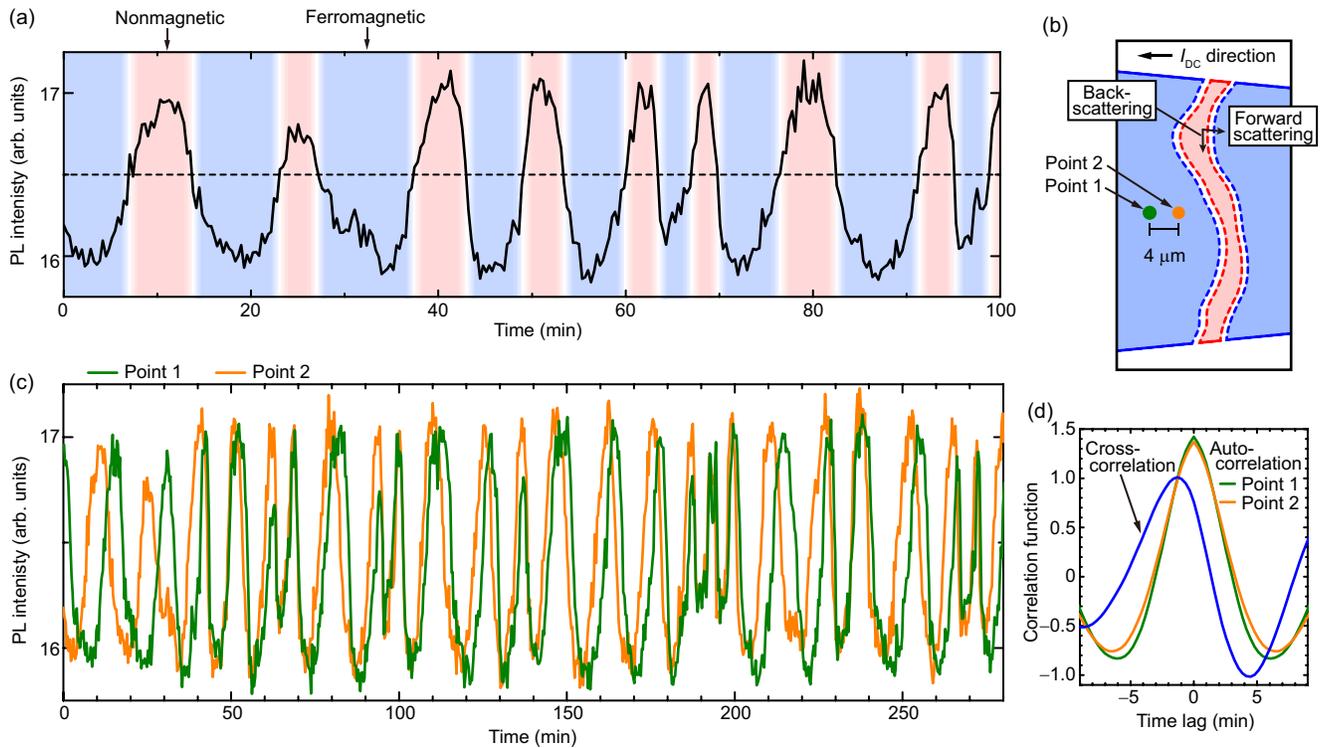} \centering\caption{(Color online) (a) $\mu$-PL
intensity at fixed point as function of time with arbitrary $t=0$,
$I_{\text{DC}}=110$~nA; $\nu=0.662$. Horizontal dotted line corresponds to
intensity when $I_{\text{DC}}=0$~nA. (b) Schematic of sample describing
locations of Points $1$ and $2$. Solid and dotted lines denote edge channels
and backscattering paths, respectively. (c) $\mu$-PL intensity as function of
time at Points $1$ (green) and $2$ (orange) in the same conditions as
Fig.~3(a). (d) Auto-correlation functions of PL intensity at Points~1 (green)
and 2 (orange), and cross-correlation function between PL intensity at
Points~1 and 2 (blue) as function of time lag. }%
\label{fig:fig3}%
\end{figure*}

To investigate the influence of nuclear spins on the domain motion, we
depolarized nuclear spins by applying r.f. radiation through a two-turn coil
wrapped around the sample. $\upsilon_{\text{domain}}$ increases when r.f. is
applied to resonantly depolarize the $^{75}$As nuclei that have been polarized
by $I_{\text{DC}}$ (Fig.~4). We applied r.f. over a range of powers, both
resonantly (red) and off-resonantly (blue). The resonance and off-resonance
frequencies were determined from the optically detected NMR spectrum taken in
this sample at the same $B$ \cite{moore}. There is a velocity difference of
$30-40$~nm/s between the two frequency cases, independent of r.f. power,
$P_{\text{r.f.}}$ (Fig.~4), which clearly indicates that polarized nuclear
spins hamper domain motion and that $\upsilon_{\text{domain}}$ can be
increased by \textit{resonantly} decreasing $P_{\text{N}}$. $\upsilon
_{\text{domain}}$ for both cases increases monotonically with $P_{\text{r.f.}%
}$. We attribute this increase to the temperature increase (Fig.~4 inset),
which also has the tendency to reduce $P_{\text{N}}$ via the thermal energy
($k_{B}T\sim5.2~\mu$eV for $60$~mK, where $k_{B}$ is Boltzman constant). This
energy can \textit{non-resonantly} decrease $P_{\text{N}}$ because of the
small Zeeman energy of nuclear spins ($\sim0.2~\mu$eV and $\sim0.4$ for
$^{75}$As and $^{71}$Ga, respectively, at $B=6.8$~T).

$\upsilon_{\text{domain}}$ is also a function of $\nu$ [Fig.~$5$(a)].
$\upsilon_{\text{domain}}$ is smallest near the phase transition ($\nu=2/3$)
and is increased by detuning $\nu$ away from $2/3$. Given that $P_{\text{N}}$
tends to slow down the propagation, the minimum in $\upsilon_{\text{domain}}$
seen near to the phase transition is expected because $P_{\text{N}}$ is
generated most effectively near the phase transition. $I_{\text{DC}}$ also
influences $\upsilon_{\text{domain}}$ [Fig.~$5$(b)]. The tendency for
monotonic decrease in $\upsilon_{\text{domain}}$ with increasing
$I_{\text{DC}}$ can be explained by the increase in $P_{\text{N}}$ generated
by the current. Later, we will also discuss other possible mechanisms which
may account for this behavior.

The low speed of these domains is noteworthy. Under the alternating current
condition used for Fig.~1(b), the current direction alternates with a $77$-ms
period, and the domain propagation length for one half-cycle is order
estimated to be $1\sim10$~nm. This is negligibly small compared to the domain
size; thus, the images under alternating current here appear static
[Fig.~1(b)] \cite{moore}.

Comparison of $\upsilon_{\text{domain}}$ to the velocity of the current is
important. The velocity of edge current $\upsilon_{\text{edge}}$
[\textit{along} the \textit{solid} blue and red lines in Fig.~3(b)] is
order-estimated to be $\upsilon_{\text{edge}}\sim\frac{I_{\text{DC}}}%
{en_{e}\ell_{B}}\sim10^{5}$~m/s, where $n_{e}\sim10^{11}$~cm$^{-2}$, $e$ is
the elementary charge, and $\ell_{B}$ is the magnetic length. Electrons
contributing to $I_{\text{DC}}$ must pass as charge current across the domain
walls bridging the two sides of the Hall bar, \textit{i.e.} forward scattering
at the domain walls. The average velocity $\upsilon_{\text{forward}}$ of the
forward scattering [\textit{across} the \textit{dotted} blue and red lines in
Fig.~3(b)], \textit{i.e.} the charge current across a domain wall, is roughly
$\upsilon_{\text{forward}}\sim\frac{\ell_{B}}{W}\upsilon_{\text{edge}}%
\sim10^{2}$~m/s, where $W$ ($\sim60$~$\mu$m) is the width of the Hall bar,
assuming a uniform current distribution over the domain wall \cite{vforward}.
$\upsilon_{\text{domain}}$ ($\sim10^{-7}$~m/s) is, therefore, $>9$ orders of
magnitude slower than the velocity of the charge current ($\sim10^{2}$~m/s).
This indicates that the domain propagation does not assist in the charge
transport of the source-drain current. Also, the propagation direction is
opposite to that caused by the spin-torque transfer mechanism. Thus, an
alternative interaction must be the cause of the domain wall motion.
Spin-torque transfer cannot be ruled out, however, as a possible explanation
of the decrease in $\upsilon_{\text{domain}}$ with increasing $I_{\text{DC}}$
[Fig. 5(b)], as this may reflect spin-torque transfer trying to move the
domains in the opposite direction to the observed propagation.

Though we cannot be certain about the specific causes of the domain
propagation, we offer here some considerations. One mechanism which might seem
to provide an intuitive explanation involves the $P_{\text{N}}$ generated by
the current as the driving force \cite{flipflop}. $P_{\text{N}}$ modifies the
local electron spin splitting energy \cite{energy}, and as a result, both
magnetic phases become more energetically favorable in the regions along the
side of the domain walls where $P_{\text{N}}$ is generated, i.e. the side
\textquotedblleft downstream\textquotedblright\ of electron flow. This creates
a local perturbation of the domain wall as electrons inside join the favorable
phase, displacing the interface. A continued cycle of $P_{\text{N}}$
generation and interface displacement causes an effective motion of the domain
walls in the upstream direction, as observed. However, this mechanism of
motion is contradicted by the observation of $P_{\text{N}}$ slowing down the
domain velocity [Fig. $4$].

\begin{figure}[t]
\begin{center}
\includegraphics{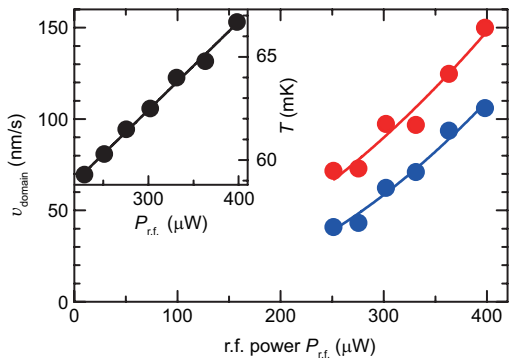}
\end{center}
\caption{(Color online) Average domain velocity in $I_{\text{DC}}$ direction
$\upsilon_{\text{domain}}$ as function of on-resonant (red, $49.7717$ MHz) and
off-resonant (blue, $49.84$~MHz) r.f. power $P_{\text{r.f.}}$; $I_{\text{DC}%
}=110$~nA, $\nu=0.664$. Inset: The dilution refrigerator mixing chamber
temperature $T$ as function of $P_{\text{r.f.}}$}%
\label{fig:fig4}%
\end{figure}

The domain motion can also be accounted for by steps in the electrochemical
potential which are formed by the backscattering channels located along the
domain walls. Electronic spin states located on the upper step along each
boundary are unstable and may reduce their potential energy by flipping their
spins to join the adjacent spin phase. Since no electrons are transported in
this process, the domain wall is displaced in the upstream direction. As $\nu$
is moved away from the phase transition, states in the domain walls become
less stable owing to the larger energy gap between the phases \cite{shibata},
and electron spins may flip more readily, causing $\upsilon_{\text{domain}}$
to grow away from the transition as observed in Fig. $5$(a).

The decrease in $\upsilon_{\text{domain}}$ with $P_{\text{N}}$ can be
accounted for, but it requires a mechanism in which $P_{\text{N}}$ is able to
diffuse across the phase boundaries, which is a process that is thought to be
inhibited by electronic spin states making up the domain walls. Because of its
direction of polarization, $P_{\text{N}}$ that diffuses across the boundaries
acts to decrease the number of nuclei available for electron spin flip-flop
exchange processes. Thus, electron spins on the upstream side of domain walls
will flip less frequently, and the domain wall motion will be slowed.

\begin{figure}[t]
\begin{center}
\includegraphics{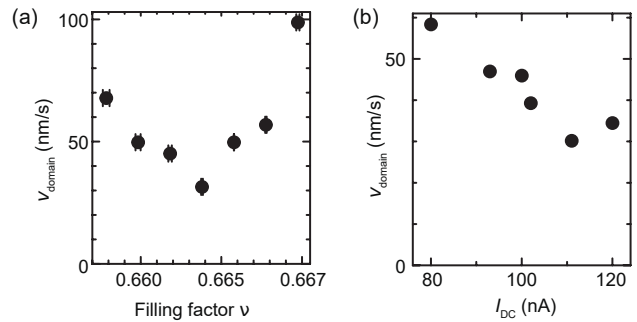}
\end{center}
\caption{(a) $\upsilon_{\text{domain}}$ as function of $\nu$; $I_{\text{DC}%
}=110$~nA. Error in $\nu$ calculated from uncertainty in electron density. (b)
$\upsilon_{\text{domain}}$ as function of $I_{\text{DC}}$; $\nu=0.664$.}%
\label{fig:fig5}%
\end{figure}

\begin{acknowledgments}
The authors are grateful to N. Shibata, K. Muraki, and T. Fujisawa for
discussions, and to Y. Hirayama and M. Matsuura for experimental support. This
work was supported by the Mitsubishi Foundation and a Grant-in-Aid for
Scientific Research (no. 24241039) from the Ministry of Education, Culture,
Sports, Science, and Technology (MEXT), Japan. J.N.M. was supported by a
Grant-in-Aid from MEXT and the Marubun Research Promotion Foundation. J. H.
was supported by a Grant in-Aid from the Tohoku University International
Advanced Research and Education Organization.
\end{acknowledgments}


\begin{thebibliography}{99}                                                                                               %


\bibitem {allwood}D. A. Allwood, G. Xiong, C. C. Faulkner, D. Atkinson, D.
Petit, and R. P. Cowburn, Science \textbf{309}, 1688 (2005).

\bibitem {parkin}S. S. P. Parkin, M. Hayashi, and L. Thomas, Science
\textbf{320}, 190 (2008).

\bibitem {yamanouchi}M. Yamanouchi, D. Chiba, F. Matsukura, and H. Ohno,
Nature \textbf{428}, 539 (2004).

\bibitem {yamaguchi}A. Yamaguchi, T. Ono, S. Nasu, K. Miyake, K. Mibu and T.
Shinjo, Phys. Rev. Lett. \textbf{92}, 077205 (2004).

\bibitem {miron}I. M. Miron \textit{et al}., Nat. Mater. \textbf{10}, 419 (2011).

\bibitem {emori}S. Emori, U. Bauer, S.-M. Ahn, E. Martinez and G. S. D. Beach,
Nat. Mater. \textbf{12}, 611 (2013).

\bibitem {franken}J. H. Franken, M. Herps, H. J. M. Swagten, and B. Koopmans,
Sci. Rep. \textbf{4}, 5248 (2014).

\bibitem {ramsay}A. J. Ramsay, P. E. Roy, J. A. Haigh, R. M. Otxoa, A. C.
Irvine, T. Janda, R. P. Campion, B. L. Gallagher and J. Wunderlich, Phys. Rev.
Lett. \textbf{114}, 067202 (2015).

\bibitem {feher}G. Feher, Phys. Rev. Lett. \textbf{3}, 135 (1959).

\bibitem {kronmuller99}S. Kronm\"{u}ller, W. Dietsche, K. von Klitzing, G.
Denninger, W. Wegscheider, and M. Bichler, Phys. Rev. Lett. \textbf{82}, 4070 (1999).

\bibitem {machida}T. Machida, T. Yamazaki, K. Ikushima, and S. Komiyama, Appl.
Phys. Lett. \textbf{82}, 409 (2003).

\bibitem {yusaN}G. Yusa, K. Muraki, K. Takashina, K. Hashimoto, and Y.
Hirayama, Nature \textbf{434}, 1001 (2005).

\bibitem {petta}J. R. Petta, A. C. Johnson, J. M. Taylor, E. A. Laird, A.
Yacoby, M. D. Lukin, C. M. Marcus, M. P. Hanson, and A. C. Gossard, Science
\textbf{309}, 2180 (2005).

\bibitem {koppens}F. H. L. Koppens, J. A. Folk, J. M. Elzerman, R. Hanson, L.
H. Willems van Beveren, I. T. Vink, H. P. Tranitz, W. Wegscheider, L. P.
Kouwenhoven, and L. M. K. Vandersypen, Science \textbf{309}, 1346 (2005).

\bibitem {ono}K. Ono and S. Tarucha, Phys. Rev. Lett. \textbf{92}, 256803 (2004).

\bibitem {yusa}G. Yusa, K. Hashimoto, K. Muraki, T. Saku, and Y. Hirayama,
Phys. Rev. B \textbf{69}, 161302(R) (2004).

\bibitem {latta}C. Latta \textit{et al}., Nat. Phys. \textbf{5}, 758 (2009).

\bibitem {vink}I. T. Vink, K. C. Nowack, F. H. L. Koppens, J. Danon, Y. V.
Nazarov, and L. M. K. Vandersypen, Nat. Phys. \textbf{5}, 764 (2009).

\bibitem {hennel}S. Hennel \textit{et al}., Phys. Rev. Lett. \textbf{116},
136804 (2016).

\bibitem {sanada}H. Sanada, Y. Kondo, S. Matsuzaka, K. Morita, C. Y. Hu, Y.
Ohno, and H. Ohno, Phys. Rev. Lett. \textbf{96}, 067602 (2006).

\bibitem {tsui}D. C. Tsui, H. L. Stormer, and A. C. Gossard, Phys. Rev. Lett.
\textbf{48}, 1559 (1982).

\bibitem {moore}J. N. Moore, J. Hayakawa, T. Mano, T. Noda, and G. Yusa,
arXiv:cond-mat http://arxiv.org/abs/1606.06416 (2016).

\bibitem {SI}See Supplemental Video on domain propagation in $6\times6$-$\mu
$m$^{2}$ PL-intensity images from an arbitrary time $\tau=0$ to $204$~min.
$I_{\text{DC}}=110$~nA, $\nu=0.664$. Image step size: $1~\mu$m. Some of the
frames of this video are displayed in Fig. 2.

\bibitem {hayakawa}J. Hayakawa, K. Muraki, and G. Yusa, Nat. Nano. \textbf{8},
31 (2013).

\bibitem {yusaPRL}G. Yusa, H. Shtrikman, and I. Bar-Joseph, Phys. Rev. Lett.
\textbf{87}, 216402 (2001).

\bibitem {wojs06}A. W\'{o}js, A. G\l adysiewicz, and J. J. Quinn, Phys. Rev. B
\textbf{73}, 235338 (2006).

\bibitem {vforward}$\upsilon_{\text{edge}}$ is the velocity through a
cross-section $\sim\ell_{B}$, whereas $\upsilon_{\text{forward}}$ is the
velocity through the cross-section $\sim W$. Thus, $\upsilon_{\text{forward}}$
is $\frac{\ell_{B}}{W}$ times smaller than $\upsilon_{\text{edge}}$, because
of the charge conservation law.

\bibitem {flipflop}Flip-flop scattering produces $P_{\text{N}}$ pointing
downward (upward) with respect to $B$ after electrons participating in the
direct current cross domain walls and join the ferromagnetic (non-magnetic) phase.

\bibitem {energy}$E_{\text{S}}=|g^{\ast}|\mu_{\text{B}}B-A\langle I_{z}%
\rangle$, where $g^{\ast}$ is the effective \textit{g}-factor of electrons,
$\mu_{\text{B}}$ is the Bohr magneton, $A>0$ is the hyperfine constant, and
$\langle I_{z}\rangle$ is the average of the nuclear spin quantum number.

\bibitem {shibata}N. Shibata and K. Nomura, Phys. Soc. Jpn. \textbf{76},
103711 (2007).
\end{thebibliography}
\end{document}